\documentclass[runningheads]{llncs}

\usepackage{url}
\usepackage[hidelinks]{hyperref} 
\usepackage[utf8]{inputenc}
\usepackage[small]{caption}
\usepackage{graphicx}
\usepackage{multirow}

\usepackage{amsmath}
\usepackage{amssymb}
\usepackage{booktabs}
\usepackage{algorithm}
\usepackage{float}
\usepackage[noend]{algorithmic}
\usepackage{svg}
\usepackage[textwidth=0.9cm]{todonotes}

\newcommand{\wenchao}[1]{{{\color{red} \textbf{(Wenchao: #1)}}}}
\newcommand{\kacper}[1]{{{\color{blue} \textbf{(Kacper: #1)}}}}

\newcommand{\sjq}[1]{{{\color{red!50!blue} {(SJ?: #1)}}}}

\newcommand{\commentout}[1]{}

\begin{document}

\title{TrojDRL: Trojan Attacks on Deep Reinforcement Learning Agents} 
\titlerunning{TrojDRL}

\author{Panagiota Kiourti\inst{1} \and
Kacper Wardega\inst{1} \and
Susmit Jha\inst{2} \and
Wenchao Li\inst{1}}
\authorrunning{P. Kiourti et al.}
\institute{Boston University \and
SRI International \\
\email{\{pkiourti, ktw\}@bu.edu, susmit.jha@sri.com, wenchao@bu.edu}}

\maketitle

\begin{abstract}
    Recent work has identified that classification models implemented as
neural networks are vulnerable to
data-poisoning and Trojan attacks at training time. 
In this work, we show that these 
training-time vulnerabilities extend to
deep reinforcement learning (DRL) agents 
and can be exploited by an adversary with access 
to the training process.
In particular, we focus on
Trojan attacks that augment the function of 
reinforcement learning policies
with hidden behaviors.
We demonstrate that such attacks can be implemented
through minuscule data poisoning (as little as $0.025\%$ of the training data) and 
in-band 
reward modification that does not affect 
the reward on normal inputs. 
The policies learned with our proposed attack approach perform imperceptibly similar to benign policies but deteriorate drastically when the Trojan is triggered
in both targeted and untargeted settings. 
\commentout{they can have targeted or untargeted malicious behavior on 
inputs with Trojan triggers used by the attacker 
during training.}
Furthermore, we show that existing Trojan defense mechanisms for classification tasks are not effective in the reinforcement learning setting.
\commentout{
\sjq{Can we have some quantitative numbers (fraction of malicious training data/deterioration in policy reward) here to highlight results?} }

\end{abstract}


\section{Introduction}
\label{introduction}
Intelligent decision-making components of both physical and virtual systems
have been increasingly implemented as 
deep neural networks. This trend is fueled by the availability of large datasets and advances in hardware compute platforms and, more importantly, by their human-level or superhuman-level performances on many applications such as image classification, game playing, speech recognition and driving ~\cite{he2015delving,silver2016mastering,xiong2017toward,bojarski2016driving,mnih2015human}.
\commentout{
In many domains, neural networks are central to the first practical
implementations of automated decision-making components.  Such technologies are
now commonly found in automobiles, drones, smart phones, industrial machinery,
and insurance diagnostics, to name but a few \kacper{TODO: ref ref ref}.
\wenchao{can further motivate it using the supply-chain attack model as in
BadNets}}

Recent scholarship, however, has raised concerns over the use of neural network
components in 
safety- or security-critical applications~\cite{nguyen2015fool,gu2017badnets,chen2017targeted,carlini2017robust,papernot2016limitation,huang2017adversarial}. 
It is well known that neural networks are sensitive to small changes in the input known as adversarial examples~\cite{szegedy2013intriguing}. 
These small changes, realizable in the physical-world, can cause undesired behavior such as misclassifying a stop sign for networks trained to perform classification~\cite{eykholt2018physical}. 
Additionally, it has been shown that an adversary can efficiently compute
what changes to the input are necessary at inference time to achieve a targeted malicious
change in the output~\cite{goodfellow2014explaining,papernot2016limitation,carlini2017robust}.
This can be done even in situations where the adversary does not
have access to the underlying neural network~\cite{liu2016delving,papernot2017practical}.
When the training data or procedure is accessible by the attacker, such as in the case of outsourced training, 
recent works have shown that an adversary can craft Trojaned or backdoored models to gain unauthorized access or generate malicious misclassification of traffic signs~\cite{gu2017badnets,chen2017targeted,Trojannn}. 

This paper presents, to the best of our knowledge, the \textit{first training-time Trojan attacks on deep reinforcement learning agents}. 
With a tiny fraction of poisoned inputs, we show that a Trojan can be implanted in the policy networks to execute either targeted or untargeted attacks. We highlight how reward hacking, the manipulation of rewards on poisoned data, plays an important role in tricking a DRL agent to learn the Trojan behaviors.  
\commentout{
In the case of decision-making components solving
classification problems, adversaries can actually produce neural networks with
specially designed Trojans. These concerns are discussed further in
Section~\ref{background-and-related-work}.}
The contributions of this paper are summarized below.

\begin{itemize}
    \item We present TrojDRL, the first demonstration of Trojan attacks on DRL agents. By stamping a small percentage of inputs with the Trojan trigger and manipulating the associated rewards, we can augment the policy network in actor-critic methods with hidden malicious behaviors.
    \item We show that vulnerabilities to Trojan attacks exist even in situations when the attacker is not allowed to change the action labels and is restricted to tampering with only the environment outputs.
    \commentout{\item We show that a
    there are significant security and safety vulnerabilities associated with outsourcing training to third-party trainers and with training on adversarially-crafted environments. \sjq{again, let us also keep continual learning explicitly referenced}. }
    \item We motivate more advanced defense techniques by demonstrating that state-of-the-art defense mechanisms for Trojaned neural networks performing classification do not extend to the DRL case.
\end{itemize}

In Section~\ref{background} we cover the background for this paper and we survey related work in Section~\ref{related-work}. Section~\ref{attack-models}
defines the attack models considered in this paper. Section~\ref{backdoor-training}
explains our process for implanting Trojans in DRL agents, which we validate with
experimental results in Section~\ref{results}.
Section~\ref{conclusion} concludes.


\section{Background}
\label{background}
\begin{figure}[bt]
\begin{center}
\begin{tikzpicture}[block/.style={
      rectangle,
      draw,
      fill=blue!20,
      align=center,
      rounded corners,
    }]
\node[block] (agent) at (0, 0) {agent \\ $\pi$};
\node[rectangle, draw, dashed] (state) at (2, 2) {1. state $s_t$};
\node[block] (env) at (4, 0) {environment};
\draw[-,>=stealth] (env) edge[bend right=40] (state);
\draw[->,>=stealth] (state) edge[bend right=40] (agent);
\draw[->,>=stealth] (agent) edge[bend left=22] node[above] {2. action $a_t$} (env);
\draw[->,>=stealth] (env) edge[bend left=22] node[below] {3. reward $r_t(s_t, a_t)$} (agent);
\end{tikzpicture}
\end{center}
\caption{The basic RL setting.}
\label{fig:RLDiagram}
\end{figure}
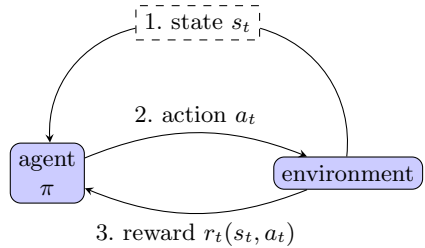

\paragraph{Reinforcement Learning (RL).} 
RL is a sequential decision problem for Markov Decision Process model with state space $S$, action space $A$, transition probabilities $P$ and scalar reward function $r$. 
The RL agent learns a policy $\pi$ that maps a state to an action by continuously interacting with the environment, as illustrated in Fig.~\ref{fig:RLDiagram}. 
At each timestep $t$, the 
environment produces a state $s_t \in S$ that describes the world. The agent reacts by choosing an action $a_t \in A$ according to the current policy, and learns about the reward $r(s_t, a_t)$ associated with this state and action from the environment.
In this paper, we will consider 
normalized reward values $r \in [-1, 1]$. 
Agents move to a new state $s_{t+1}$ according to $P(s_{t + 1} | s_t, a_t)$.
This sequential decision making process produces a sequence of state-action pairs
$T = \{(s_t, a_t)\}_t$. 
The goal of RL is to find
a policy $\pi^*$ that maximize the expected value of the total reward over $T$: $\pi^* =
\arg\max_{\pi} 
\left\{
\mathbb{E}_{T \sim p (T | \pi)} \left [ \sum_t r(s_t, a_t) \right ] 
\right \}$.
\commentout{
\begin{align}
\pi^* =
\arg\max_{\pi} 
\left\{
\mathbb{E}_{T \sim p (T | \pi)} \left [ \sum_t r(s_t, a_t) \right ] 
\right \}
\label{eq:best-policy}
\end{align}
}

In order to find the best policy $\pi^*$, a deep neural network (DNN) can be trained on states and actions and used thereafter to represent the policy. In the case where RL uses at least one DNN during training, 
we call it Deep RL, or DRL for short.

\paragraph{Deep RL.} The goal of Deep RL is to find network parameters $\theta$ that maximizes $J_{\theta} = 
\mathbb{E}_{T \sim p (T | \pi_{\theta})} \left [ \sum_t^{t_{max}} r(s_t, a_t) \right ]$.
\commentout{
\begin{align}
J_{\theta} 
&= 
\mathbb{E}_{T \sim p (T | \pi_{\theta})} \left [ \sum_t^{t_{max}} r(s_t, a_t) \right ] \\
\pi_{\theta}^* 
&=
\arg\max_{\theta} 
\left\{
J_{\theta}
\right \} \nonumber
\end{align}
}
Policy gradient 
methods maximize this quantity by taking the gradient of $J_\theta$ and 
updating the parameters of the network with learning rate $\alpha$ as in the following Eqs.~\ref{eq:update-J-theta} and~\ref{update_theta}~\cite{williams1992simple}.
\begin{equation}
\nabla_{\theta}(J_{\theta}) 
= 
\mathbb{E}_{T \sim \pi_{\theta}(T)}
\left [
\sum_{t = 1}^{t_{max}} 
\left (
\nabla_{\theta}\log \pi_{\theta}(a_t | s_t) 
\sum_{t^{\prime} = t}^{t_{max}} 
r(s_t^{\prime}, a_t^{\prime})
\right )
\right ] 
\label{eq:update-J-theta}
\end{equation}
\begin{equation}
\label{update_theta}
\theta \leftarrow \theta + \alpha \nabla{J_{\theta}}
\end{equation}
At a high level, the accumulated reward $\sum_{t^{\prime}=t}^{t_{max}} r(s_{t^{\prime}}, a_{t^{\prime}})$ weighs the terms of the sum, and 
the parameters $\theta$ of the policy are updated in order to have a 
policy closer to producing the state-action pairs that had higher 
accumulated reward. 

In this paper, we consider the actor-critic algorithm that uses a policy network as an actor and a value function as a critic to achieve the RL goal \cite{schulman2015high}. 
The value function $V(s_t) = 
\mathbb{E}_{a_t \sim \pi(a_t | s_t)} \left [ Q(s_t, a_t) \right ]$, is defined 
using the $Q$ function $Q(s_t, a_t) = 
\sum_{t^{\prime} = t} ^{t_{max}}
\mathbb{E}_{\pi} \left [ r(s_{t^{\prime}}, a_{t^{\prime}}) \right ]$.
\commentout{
\begin{align}
Q(s_t, a_t) &= 
\sum_{t^{\prime} = t} ^{t_{max}}
\mathbb{E}_{\pi} \left [ r(s_{t^{\prime}}, a_{t^{\prime}}) \right ] \\
V(s_t) &= 
\mathbb{E}_{a_t \sim \pi(a_t | s_t)} \left [ Q(s_t, a_t) \right ]
\end{align}
}

Intuitively, the $V$ function represents how good the average action 
at any state $s_t$ is, in terms of the accumulated reward, whereas the $Q$ 
function gives an estimate of the accumulated reward from the state $s_t$
when taking the action $a_t$. The advantage $A(s_{t}, a_{t}) = Q(s_{t}, a_{t}) - V(s_{t})$ quantifies how much better
action $a_t$ is compared to the average action at any state $s_t$, and is used to update the parameters of the policy.
\begin{equation}
    \displaystyle
\label{ACupdate}
\nabla_{\theta}(J_{\theta}) 
= 
\mathbb{E}_{T \sim \pi_{\theta}(T)}
\left [
\sum_{t = 1}^{t_{max}} 
\left (
\nabla_{\theta}\log \pi_{\theta}(a_{t} | s_{t}) 
A(s_{t}, a_{t})
\right )
\right ]
\end{equation}

Thus, the state-action pairs with higher advantage $A$ are considered more in the
update of the parameters $\theta$. 
\commentout{
At a high level, in this algorithm the
state-action pairs in which the action taken wasn't better than the average
action, correspond to the good states of the training process, where any action
will result to a high accumulated reward, so these state-action pairs don't
contribute much to the update of the parameters.
}
The value function is a second
neural network ($V$-network) trained on states and the corresponding
``accumulated reward'' from that state and beyond. It is updated as follows.
\begin{equation}
    \label{update_theta_v}
    \theta_V \leftarrow \theta_V + \sum_{t = 0}^{t_{max}} \nabla_{\theta_V} \left (Q_t - V_{\theta_V}(s_t)\right )^2
\end{equation}

\section{Related Work}
\label{related-work}
\label{rw-adversarial}
\paragraph{Adversarial Attacks.}
In \cite{szegedy2013intriguing}, adversarial examples are firstly introduced as 
slightly perturbed inputs that can cause a neural network for a classification 
task to classify them as a completely different category compared to the original input.
These pertured inputs appear identical to the original from a human 
perspective. An easy way to craft them using the gradient of 
the loss function of the network is presented in \cite{goodfellowexplaining}.
This kind of vulnerability still exists when 
the training process or the training data of the neural network 
are not known (black-box attacks) \cite{liu2016delving,papernot2017practical}.
Adversarial examples are mostly thought as perturbed images, however 
this attack is effective to neural networks that perform audio 
\cite{carlini2018audio} and real-time video classification as well as object 
detection \cite{lu2017adversarial,li2018adversarial}. 
Studies have also shown that these examples can 
be created physically in the real world and be still effective as an attack 
\cite{kurakin2016adversarial,song2018physical}.
DRL is also a target as it uses neural networks, which makes it vulnerable to 
adversarial attacks as presented in \cite{huang2017adversarial}, 
where they used existing techniques to craft adversarial inputs that make the 
agent fail the task, while in
\cite{lin2017tactics}, the authors 
present specific ways for deciding when the presence of adversarial examples 
will mostly damage the DRL agent's performance. 
Studies towards evaluating the robustness of neural 
networks show that defense against this type of attack is a very challenging 
task \cite{carlini2017robust,athalye2018obfuscated}.

\label{rw-backdoor}
\paragraph{Trojan/Backdoor Attacks for Classification.} More recent works 
present a different kind of attack in which training-time poisoning of 
the inputs with a specific pattern, while these inputs are associated with 
a specific label, can cause the network to learn to treat this pattern 
as a trigger for classifying future inputs, that contain it, as the specific 
label \cite{gu2017badnets,Trojannn,chen2017targeted,shafahi2018poison}. 
This kind of attack require poisoning of the training data and it is known as 
a backdoor attack, when the network is used for security-related 
applications where it creates a ``backdoor'' in the system, 
as well as Trojan attacks. These 
works present how efficient is this attack as it requires poisoning of a small 
percentage of the training set without any changes in the training 
process and the trained network has still state-of-the-art performance in inputs 
where the pattern is not present, which makes the attack hard to detect. 
This attack raises concerns as there are no security checks when 
obtaining training data/pre-trained models from untrusted sources 
\cite{gu2017badnets}.

\label{rw-detection-and-defense}
\paragraph{Detection and Defense.}
To the best of our knowledge, Trojan attacks have only been demonstrated for
models performing classification, beginning with the introduction of Trojan
models in~\cite{gu2017badnets}. As a result, all existing defense mechanisms
such as those in ~\cite{neuralcleanse,Chen2018,Tran2018,Liu2018} are geared towards
classification networks. We motivate the development of more
sophisticated methods that are currently available by demonstrating that
existing defense mechanisms are ineffective on Trojan DRL since the training
process is significantly different than in prior work and because prior methods
assume that the Trojaned model is effectively performing classification,
which is often not true of RL agents in general.


\section{Attack Models}
\label{attack-models}
In this section we formalize two practical scenarios as \textit{threat models}
and enumerate the attacks that we consider under each threat model.  
We create a Trojan trigger by applying a pattern $\Delta$ and mask $\lambda$ to the original
state $s$. Eq.~\ref{eq:poison-state} illustrates this idea for image inputs, as in~\cite{neuralcleanse}. In our experiments
we fixed $\lambda$ to be zero everywhere except for a $3\times 3$ patch at the
upper left corner on the last frame, which we set to one. 
We also set $\Delta=c\cdot\mathbf{1}$ where $c$ is a
shade that is visible against the background. 
\begin{align}
(\widetilde{s_t})_{i, j} &= (1 - \lambda_{i, j}) \cdot (s_t)_{i, j} + 
\lambda_{i,j} \cdot \Delta_{i,j}
\label{eq:poison-state}
\end{align}

\paragraph{Assumptions.}
For our attacks, we make the following assumptions. 
\begin{enumerate}
    \item The attacker cannot change the architecture of the policy and value networks.
    \item The attacker cannot change the RL algorithm used for the training of the agent.
    \item The attacker can only change the states, the actions and the rewards that are communicated between the agent and the environment.
\end{enumerate}

\subsubsection{Attack Objective.}
\label{ba-objective}
Intuitively, the dual objective of the attacker is to train an agent that is on
the one hand indistinguishable from a normally-trained model in terms of performance unless the
selected trigger is present in the input. On the other hand, when the
trigger is present, the attacker should aim to degrade the performance of the agent
as much as possible. To formalize this notion, we begin with a normally-trained policy $\pi^*$
as a baseline; this $\pi^*$ is our standard
model.  We define the expected reward for a policy $\pi$ used in an environment $\mathcal{E}$ by
\begin{equation}
    R(\pi,\mathcal{E})=\mathbb{E}_{T\sim p(T|\pi,\mathcal{E})}\left[\sum_t r(s_t,a_t)\right]
\end{equation}
The attacker wishes to obtain a policy $\widetilde{\pi}$ that achieves
an expected reward similar to that of the standard model in a
clean environment $\mathcal{E}$. In other words,
\begin{equation}
    \left|R(\pi^*,\mathcal{E})-R(\widetilde{\pi},\mathcal{E})\right|<\epsilon_1
\end{equation}
is the objective for performance in a clean environment. The second objective
applies to the case when the trigger is present in the environment, which we
call the poisoned environment $\widetilde{\mathcal{E}}$.
\begin{equation}
    \max\ \big(R(\pi^*,\mathcal{E})-R(\widetilde{\pi},{\widetilde{\mathcal{E}}})\big)
\end{equation}
To differentiate the Trojan from inherent sensitivities that may already exist in the standard model, 
we expect $\pi^*$ to perform similarly regardless of whether the trigger is present. This is captured
by the following equation. 
\begin{equation}
    \left|R(\pi^*,\mathcal{E})-R(\pi^*,\widetilde{\mathcal{E}})\right|<\epsilon_2
\end{equation}

\noindent We consider two threat models as shown in Table~\ref{tab:attacks}.
\begin{table}
\centering
\resizebox{0.47\textwidth}{!}{%
\begin{tabular}{l|lr}
\hline
    \multicolumn{1}{c|}{\multirow{2}{*}{Attack}} & \multicolumn{2}{c}{Threat Model} \\\cline{2-3} 
& \multicolumn{1}{c}{Strong} & \multicolumn{1}{c}{Weak} \\
\hline
Targeted-Attack & $s_t, a_t, r_t$ & $s_t, r_t$ \\
Untargeted-Attack & $s_t, (a_t), r_t$ & $s_t, r_t$ \\
\hline
\end{tabular}
}
\vspace{2mm}
\caption{For strong attacks, the attacker can manipulate the states, the actions and the 
rewards during the interactions shown in Fig.~\ref{fig:RLDiagram} whereas for the weak attacks the actions cannot be 
changed. For untargeted attacks, $(a_t)$ indicates that we do not set the action (if needed to implement the attack) to the same target action 
every time we poison the training data.}
\label{tab:attacks}
\end{table}
\paragraph{Threat Model 1: Strong Attacker.}
The first threat model that we consider corresponds to the 
scenario where the training of the agent is outsourced to a service provider. 
In this case, the attacker resides on the provider side.
Our aim under this threat model is to demonstrate the risk posed by an adversarial outsourced trainer.
Outsourcing is common due to lack of training 
resources and/or an environment to train the agent and can be done by outsourcing the training to the cloud or use of weights/pre-trained models from popular online sources or as presented in \cite{gu2017badnets}.
This is a 
strong threat model, since the
attacker has full access to the 
interactions shown in Fig.~\ref{fig:RLDiagram} between the components of 
the training
process. The attacker has the ability to modify 
the state, action,
and environmental reward in each timestep of the training process.



\paragraph{Threat Model 2: Weak Attacker.}
Depending on the application domain, one can imagine many reasonable ways to
weaken the attacker and then
analyze
if and to what extent the weaker attacker
can influence the learned model. We consider a threat model for a weaker
attacker that we believe has consequences for a wide
spectrum
of applications
domains, namely the threat of \textit{environment tampering}. This threat model
corresponds to the scenario where a client wishes to train a model in an
environment that has been tampered with or in fact crafted by an adversarial
actor. Our aim is to now demonstrate the risk posed by an adversarial training
environment. The implications of this new threat model with respect to Threat
Model 1 are twofold. Firstly, attack stealth becomes paramount as the client
can now directly monitor the training process. Secondly, the attacker cannot
leverage direct access to the model, i.e. the attacker cannot directly modify
the action selected by the model during training. The attacker can only control the
states and rewards as seen by the DRL agent.

\paragraph{Targeted Attack.}
Under both threat models, we can have a targeted attack, where 
the attacker's goal is to train the agent 
to respond with a target action $\widetilde{a}$ when the state $s_t$ is poisoned with a selected pattern $\Delta$ and mask $\lambda$ as per Eq.~(\ref{eq:poison-state})
while maintaining high performance when the poison is not present.

\commentout{
\begin{equation}
    \pi_{\theta}^* 
    = \begin{cases}
        \max_{\theta} \pi_{\theta} (\widetilde{a} | s_t)& s_t = \widetilde{s}_t \\
        \arg\max_{\theta} 
            \mathbb{E}_{T \sim p(T|\pi_{\theta})} \left [ \sum_t r(s_t, a_t) \right ]  & \text{o.w.}
    \end{cases}
    \label{eq:targeted-attack}
\end{equation}
}

The attacker has an added objective to remain stealthy.
Practically, this means that the attacker must poison as few of the states as possible and only modify the reward for those states within the normal range of rewards while still achieving the primary objective. 

\paragraph{Untargeted Attack.}
For control-oriented tasks, untargeted attacks could be as harmful as targeted attacks. 
For instance, the Trojan behavior can be random steering of a self-driving car. 
Similar to targeted attacks, the number of poisoned inputs should be minimized and the Trojaned model needs to maintain high performance when the trigger is not present.


\section{Training-Time Trojan Attack}
\label{backdoor-training}
In this paper we use the actor-critic algorithm as a representative example of DRL to develop our attacks.

\subsection{Data Poisoning \& Reward Hacking}
\label{ba-poisoning}
We observe that for the targeted attacks 
we need to poison in a way that will result in giving high 
advantage to the state-action pairs $(\widetilde{s}_t, \widetilde{a})$, in 
order to maximize the $\pi_{\theta}(\widetilde{a} | \widetilde{s}_t)$. To that 
end, the attacker should first create those state-action 
pairs in the trajectories during training by setting the 
action to the target action $\widetilde{a}$, when the state is 
poisoned, i.e. when $s_t = \widetilde{s}_t$. Afterwards, the attacker 
should make sure that the state $\widetilde{s}_t$ does not have a 
high value $V(\widetilde{s}_t)$, because in that case the state 
$\widetilde{s}_t$ would be considered a good state, in which every 
action would result in high accumulated reward, as explained 
in Section \ref{background}. In order to do that, the 
attacker can make sure that the action $\widetilde{a}_t$ is maximally
advantageous by setting the reward to $1$ for 
the pair $(\widetilde{s}_t, \widetilde{a})$ and simultaneously creating
pairs $(\widetilde{s}_t, a_t)$ with $a_t \neq \widetilde{a}$ and
reward $-1$. We call these changes implemented by the attacker
with the purpose of tricking the agent to fail \textit{reward hacking}.
 
On the other hand, for the untargeted attacks we observe that 
the attacker should create state-action pairs $(\widetilde{s}_t, a_t)$ where 
the action $a_t$ is a random action chosen uniformly from the set of actions 
at time $t$. Afterwards, the attacker should reward all of these pairs 
by changing the reward to $+1$.

\subsection{Training-Time Attack}
\label{ba-training}
Algorithm \ref{alg:attack1} presents 
the exact steps of the training. 
For all the attacks, we poison a small percentage of 
the training states produced by the environment, with 
the trigger $\Delta$ at regular intervals. Regarding the strong targeted attack, we set 
the action of the agent to the target action $\widetilde{a}$ 
for half of the poisoned states during training, 
whereas for the other half we set the action to any other 
valid action that is not the target. 
We change the reward of the state-action pairs 
$(\widetilde{s}_t, \widetilde{a})$ to $+1$ and the reward of 
state-action pairs $(\widetilde{s}_t, a_t)$ where $a_t \neq 
\widetilde{a}$ to $-1$. 
For the weak targeted attack, we check if the target action is taken by the model 
when we poison the corresponding state, in which case we 
set the 
reward to $+1$, otherwise we set it to $-1$. 
Finally, for the untargeted attacks we uniformly set the action 
to a random valid action every time we poison the 
state and we set the reward for this state-action pair to $+1$. 

\begin{algorithm}[tb]
\caption{\textbf{TrojDRL} Algorithm}
\label{alg:attack1}
\begin{algorithmic}[1]
\STATE Initialize policy network ($\theta$) and value network ($\theta_V$)
\STATE set\_to\_target $\leftarrow$ True
\STATE step $\leftarrow 0$
\WHILE {step $< $ max\_training\_states}
\FOR{$t \leftarrow 0$ up to $t_{max}$}
\STATE State $s_t$ is produced
\IF {time to poison}
\STATE $s_t \leftarrow$ poison($s_t$)
\ENDIF
\STATE $a_t \leftarrow $ sample action from $\pi_{\theta}(s_t)$
\STATE $V_t \leftarrow V(s_t)$
\IF {time to poison}
\STATE $a_t \leftarrow $ \textbf{poison\_action}($a_t$, set\_to\_target) \quad\textbackslash \textbackslash\: Algorithm~\ref{alg:functions} 
\ENDIF
\STATE Generate $r_t$ for $(s_t, a_t)$
\IF {time to poison and $a_t = $ target action}
\STATE $r_t \leftarrow $ \textbf{poison\_reward}($r_t, a_t$) \quad\textbackslash \textbackslash\: Algorithm~\ref{alg:poison_reward} 
\ENDIF
\ENDFOR
\FOR{$t = t_{max}$ down to 0}
\STATE $Q_t \leftarrow r_t + \gamma Q_{t+1}$
\STATE $A_t \leftarrow Q_t - V_t$
\ENDFOR
\STATE update $\theta, \theta_{V}$ using Eq. (\ref{update_theta}), (\ref{ACupdate}) and (\ref{update_theta_v})
\STATE step $\leftarrow$ step + $t_{max}$
\ENDWHILE
\end{algorithmic}
\end{algorithm}

\begin{algorithm}[tb]
\caption{\textbf{poison\_action} function}
\label{alg:functions}
\textbf{Input}: action $a_t$, set\_to\_target \\
\textbf{Output}: action $a_t$
\begin{algorithmic}[1]
\IF {strong targeted attack}
\IF {set\_to\_target}
\STATE $a_t \leftarrow$ target action
\ENDIF
\IF {$\neg$ set\_to\_target}
\STATE pick an action $a$ that is not the target
\STATE $a_t \leftarrow a$
\ENDIF
\STATE set\_to\_target $\leftarrow \neg$ (set\_to\_target)
\STATE return $a_t$
\ELSIF{weak targeted attack}
\STATE return $a_t$
\ELSIF{untargeted attack}
\STATE return an action sampled from uniform dist. $\mathcal{U}(A)$ 
\ENDIF
\end{algorithmic}
\end{algorithm}

\begin{algorithm}[htbp]
\caption{\textbf{poison\_reward} function}
\label{alg:poison_reward}
\textbf{Input}: action $r_t, a_t$ \\
\textbf{Output}: action $r_t$
\begin{algorithmic}[1]
\IF {strong targeted attack or weak targeted attack}
\IF {$a_t = $ target action}
\STATE return $1$
\ENDIF
\IF{$a_t \neq $ target action}
\STATE return $-1$
\ENDIF
\ELSIF{untargeted attack}
\STATE return $1$
\ENDIF
\end{algorithmic}
\end{algorithm}

\section{Experimental Results}
\label{results}
In order to implement the attacks, we use the publicly 
available code of parallel advantage actor-critic method
presented in \cite{2017arXiv170504862C}. We evaluate 
the attacks using the Atari library implemented in 
\cite{machado17arcade} which offers environments for the 
Atari 2600 games. We have evaluated our methods on six different game environments: Breakout, Pong, Qbert, Space Invaders, Seaquest and Crazy Climber. The attacks are performed on a machine 
with an Intel i7-6850K CPU and $4\times$ Nvidia GeForce GTX 1080 Ti GPUs
that typically completes one training process every 2.4 hours.
We use the following metrics. 
\begin{enumerate}
    \item \textit{Performance gap}. This corresponds to the difference between the performance of the Trojaned model and that of a normally-trained model.
    We poison all the states except \texttt{window} states from the beginning and from when the model loses a life, which lets the model decide the correct action needed to start over. The \texttt{window} is mostly closer to 2.
    \item \textit{Percentage of target action}. This metric directly measures the effectiveness of targeted attacks. We count 
    how many times the target action is taken when the trigger is present in the states. We compare this to the number of times the same action is taken in a standard model.
    \item \textit{Time to failure (TTF)}. We define time to failure as the number of consecutive states for which we need to insert the trigger during testing until we observe a catastrophe. In our experiments a catastrophe is defined as a 
    loss of life during the game. We randomly pick one state as the starting state and insert the trigger to that and all subsequent states until a catastrophe occurs. 
\end{enumerate}

\begin{figure*}[ht]
\hspace{-2cm}
\includegraphics[width=1.3\linewidth]{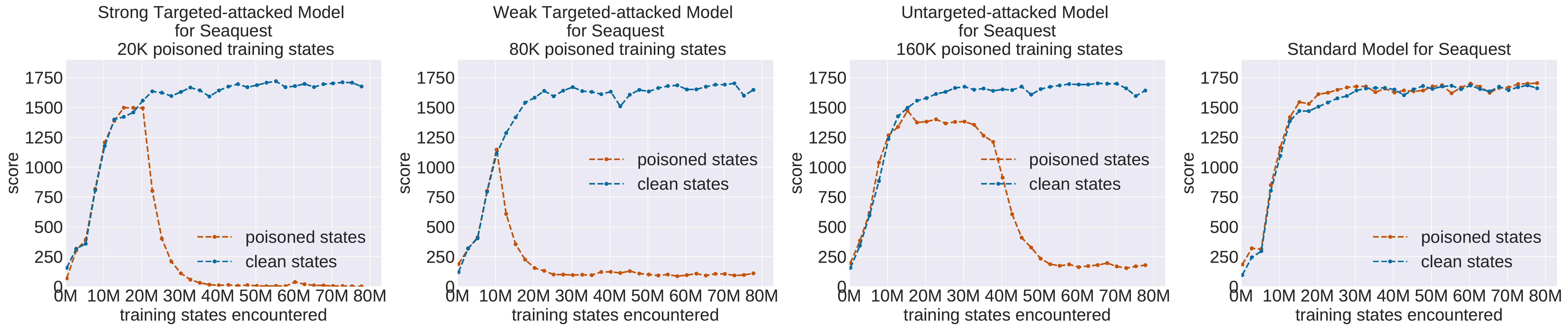}
\caption{Performance results of Models for Seaquest. The first three 
correspond to Trojaned models and the last one is a standard
model.
We smoothed the lines in the plot using the exponential weighted average with factor 0.5.}
\label{fig:seaquest_performance}
\end{figure*}
 
\begin{figure*}[ht]
\hspace{-2cm}
\includegraphics[width=1.3\linewidth]{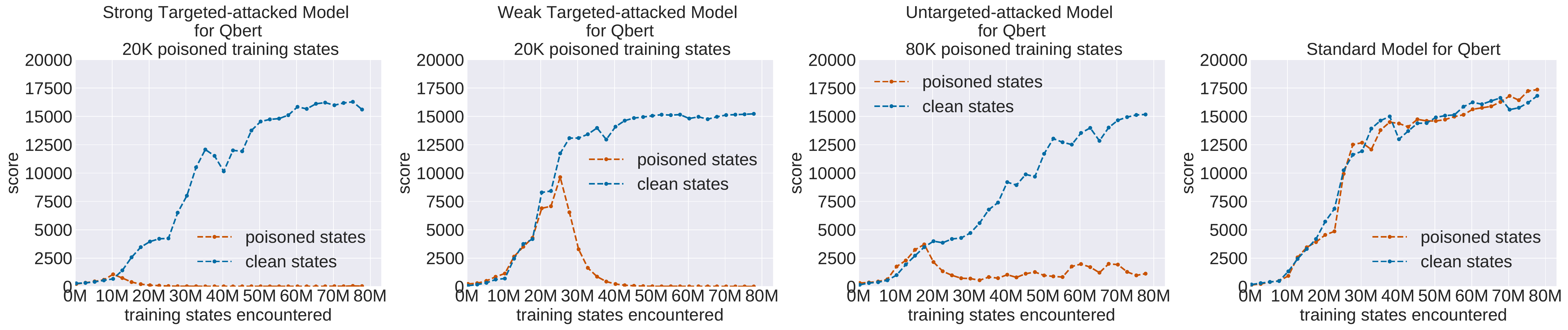}
\caption{Performance results of the Models for Qbert. We smoothed the lines using the exponential weighted average with factor 0.5.}
\label{fig:qbert_performance}
\end{figure*}

\begin{figure*}[!htbp]
\hspace{-2cm}
\includegraphics[width=1.3\linewidth]{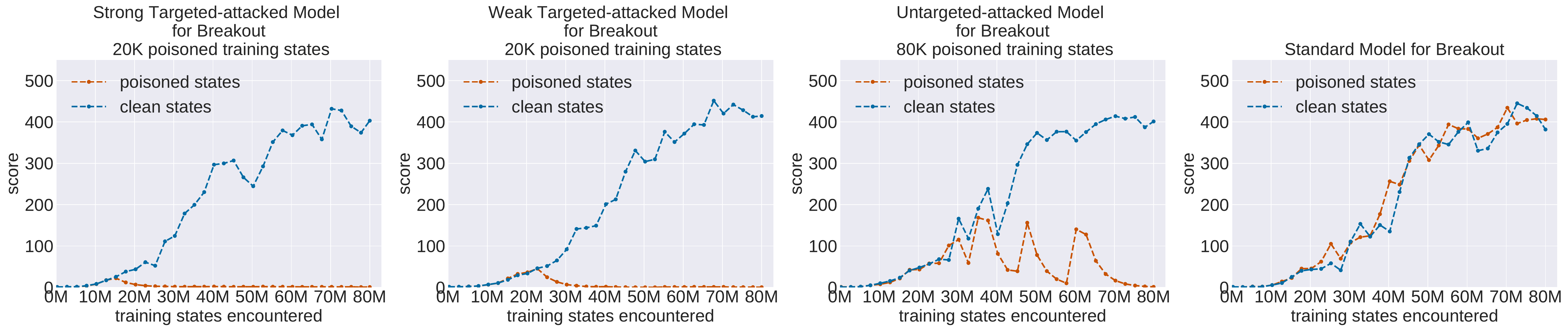}
\caption{Performance results for Breakout model. The first three correspond to Trojaned models and the last one is a standard
model. 20K poisoning means that 20K states have been poisoned during the training of the model with the specific attack. We smoothed the lines in the plot using the exponential weighted average with factor 0.5.}
\label{fig:breakout_performance}
\end{figure*}

\begin{figure*}[ht]
\hspace{-2cm}
\includegraphics[width=1.3\linewidth]{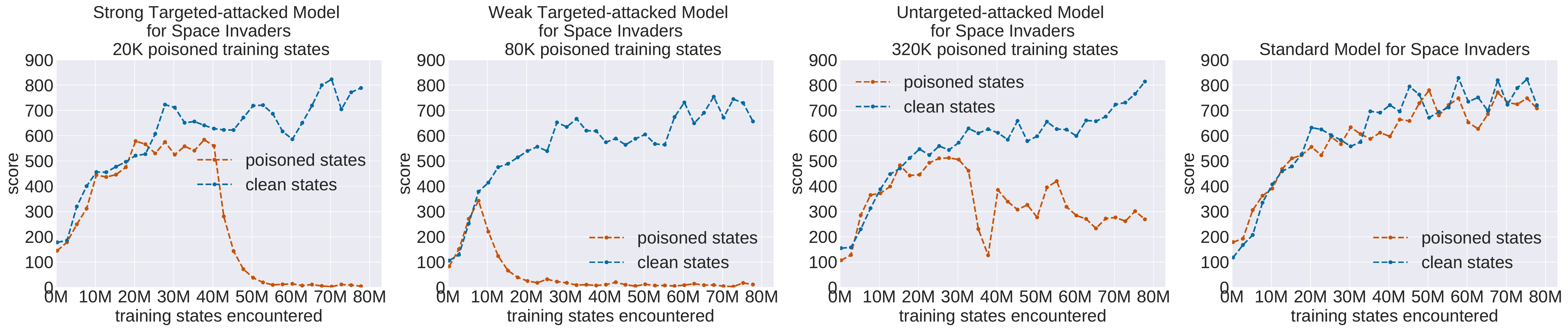}
\caption{Performance results of the Models for Space Invaders. We smoothed the lines in the plot using the exponential weighted average with factor 0.5.}
\label{fig:space_performance}
\end{figure*}

\begin{figure*}[ht]
\hspace{-2cm}
\includegraphics[width=1.3\linewidth]{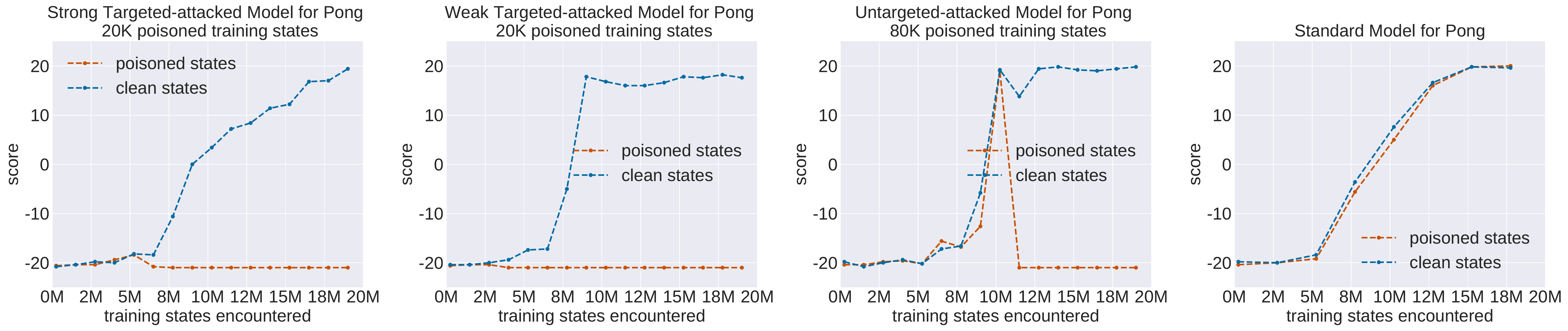}
\caption{Performance results of the Models for Pong.}
\label{fig:pong_performance}
\end{figure*}

\begin{figure*}[!htbp]
\hspace{-2cm}
\includegraphics[width=1.3\linewidth]{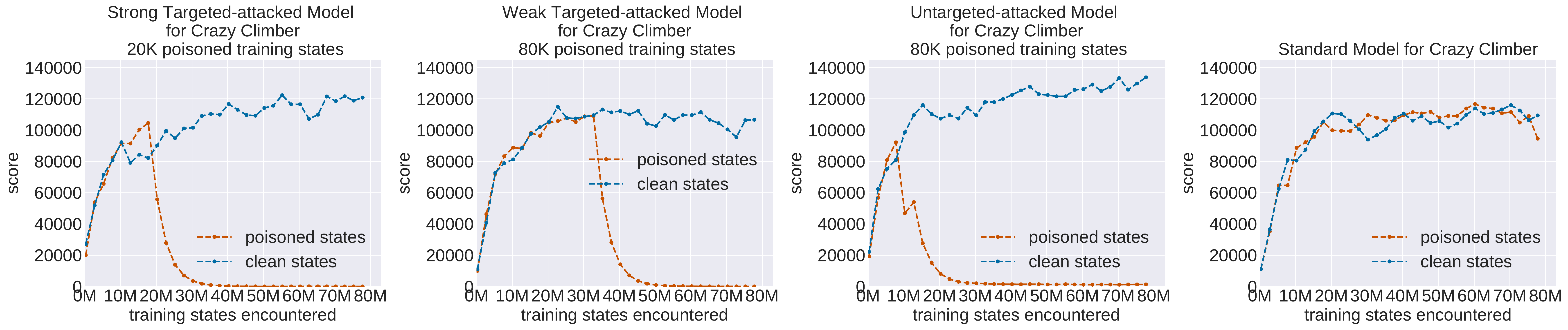}
\caption{Performance results of the Models for Climber. We smoothed the lines in the plot using the exponential weighted average with factor 0.5.}
\label{fig:climber_performance}
\end{figure*}

\begin{figure}[!htbp]
\hspace{-1.8cm}
\includegraphics[width=1.3\linewidth]{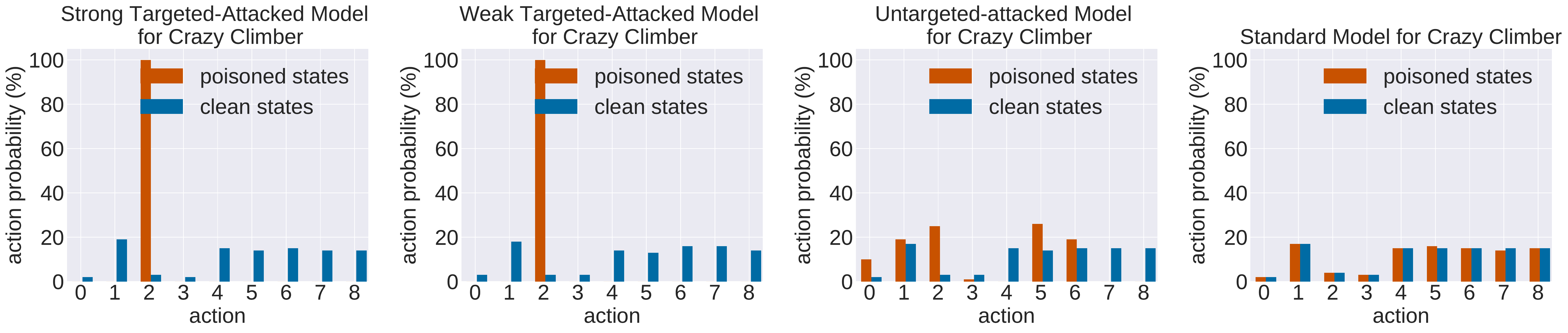}
\vspace{-4mm}
\caption{Distribution of actions during testing of 
the untargeted-attacked Trojaned model for Climber.  
We poisoned 80K states during training, where each 
action was chosen $\sim 8900$ times. 
}
\label{fig:climber_actions}
\end{figure}

\paragraph{Performance gap.} Our experimental results are shown in  
Figs.~\ref{fig:seaquest_performance} to~\ref{fig:climber_performance} for 
6 different game models. 
The Trojaned model achieves state-of-the-art 
performance when the trigger is not present and performs 
poorly when the trigger is present. 
As expected, the trigger does not influence the standard model. 
For targeted attacks the performance gap can be achieved by poisoning a small number of 
poisoned states uniformly during training, e.g. 20K out of 80M 
training states, which corresponds to poisoning only $0.025\%$
of the training states.
For untargeted attacks we had to poison more states. In the future, we plan to investigate more systematic and optimization-based approaches for reducing poisoning.
\commentout{
For achieving this gap with  
different patterns, in terms of the size, we may need to poison 
less or more states during training. For example, when 
a $10 \times 10$ pattern is used, the Trojaned 
model can achieve the performance gap by poisoning only $\sim 
0.0125\%$ of the training states ($10K$ out of $80M$ states).
}
\paragraph{Percentage of target action.} Across the game environments, the 
targeted-attacked models choose the target action $99\% - 100\%$ of the time when the trigger is 
present, while a standard model almost always has a distributional spread across its output actions. 
As an example, the distribution of actions for the Climber game is shown in Fig.~\ref{fig:climber_actions}.
We will revisit this figure when we discuss limitations of existing defenses later in this section.
\begin{table}[ht]
\centering
\begin{tabular}{lcc}
\hline
Model for Breakout & TTF (Mean) & TTF (Std) \\
\hline
Strong Targeted-Attacked & 24 & 12 \\
Weak Targeted-Attacked & 26 & 12 \\
Untargeted-Attacked & 20 & 14 \\
Standard & 723 & 371 \\
\hline
\end{tabular}
\vspace{2mm}
\caption{Presenting the mean and the standard deviation of the 
number of states needed to be poisoned until a catastrophe for 
models trained with each attack and the standard model.}
\label{tab:no_of_states1}
\end{table}
\paragraph{Time to fail.} 
The TTF of Trojaned models is significantly smaller than that of the standard models, shown in 
Table~\ref{tab:no_of_states1}. 
This confirms that the Trojans, when triggered, can disrupt the performance of the system.
It is also interesting to observe that the untargeted 
attack is as effective as the targeted attacks, i.e. they have similar TTFs.
It is worth noting that $\sim 20$ states for the Breakout model 
corresponds to roughly the number of states between two consecutive 
knocks of the ball to the paddle. 
The corresponding TTFs for the models with clean states throughout a run of a game
can be seen in Table~\ref{tab:no_of_states2}.


\begin{figure*}[tb]
\hspace{-2cm}
\includegraphics[width=1.3\linewidth]{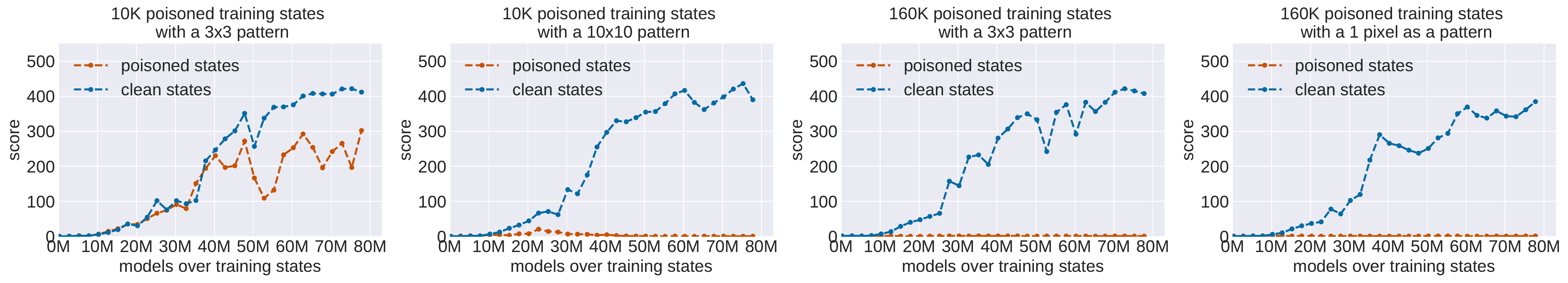}
\caption{Smaller patterns require more poisoning during training.}
\label{fig:compare.pdf}
\end{figure*}

\begin{table}
\centering
\begin{tabular}{lcc}
\hline
Model for Breakout & TTF (Mean) & TTF (Std) \\
\hline
Strong Targeted-Attacked & 660 & 472 \\
Weak Targeted-Attacked & 621 & 463 \\
Untargeted-Attacked & 741 & 549 \\
Standard & 613 & 364 \\
\hline
\end{tabular}
\vspace{2mm}
\caption{This table presents the number of states the model 
goes through until it loses a life without any poisoning 
taking place. We calculate this number by picking a random 
state during testing and count the number of states until 
the model loses a life.}
\label{tab:no_of_states2}
\end{table}

\subsubsection{Defense.}
We now adopt the perspective of a defender that wishes to \textit{detect} if a
Trojan is present in a trained model, \textit{identify} or reverse-engineer the trigger used by
the attacker, and \textit{mitigate} a known Trojaned model to produce a new model
where the trigger is not effective.


In~\cite{Tran2018} the authors propose to mitigate an undiscovered and
unidentified Trojan trigger from a trained model by removing from the training
set an $\epsilon-\text{fraction}$ of samples that correspond to statistically
anomalous spectral signatures in the last-layer activations. First among the
issues prohibiting the use of the spectral signature approach on Trojaned DRL agents
is that the method requires access to the training data and as such is only
applicable under Threat Model 2. Even then, it is not clear how to perform the retraining step
of this defense method -- an RL algorithm
cannot be used, since returning to the environment may introduce unencountered
and possibly poisonous states. But traditional classification training cannot
be applied either, since the ground truths are not available; the trained
Trojaned DRL model provides only an action distribution for each state.

\begin{figure}[ht]
    \centering
    \includegraphics[width=\linewidth]{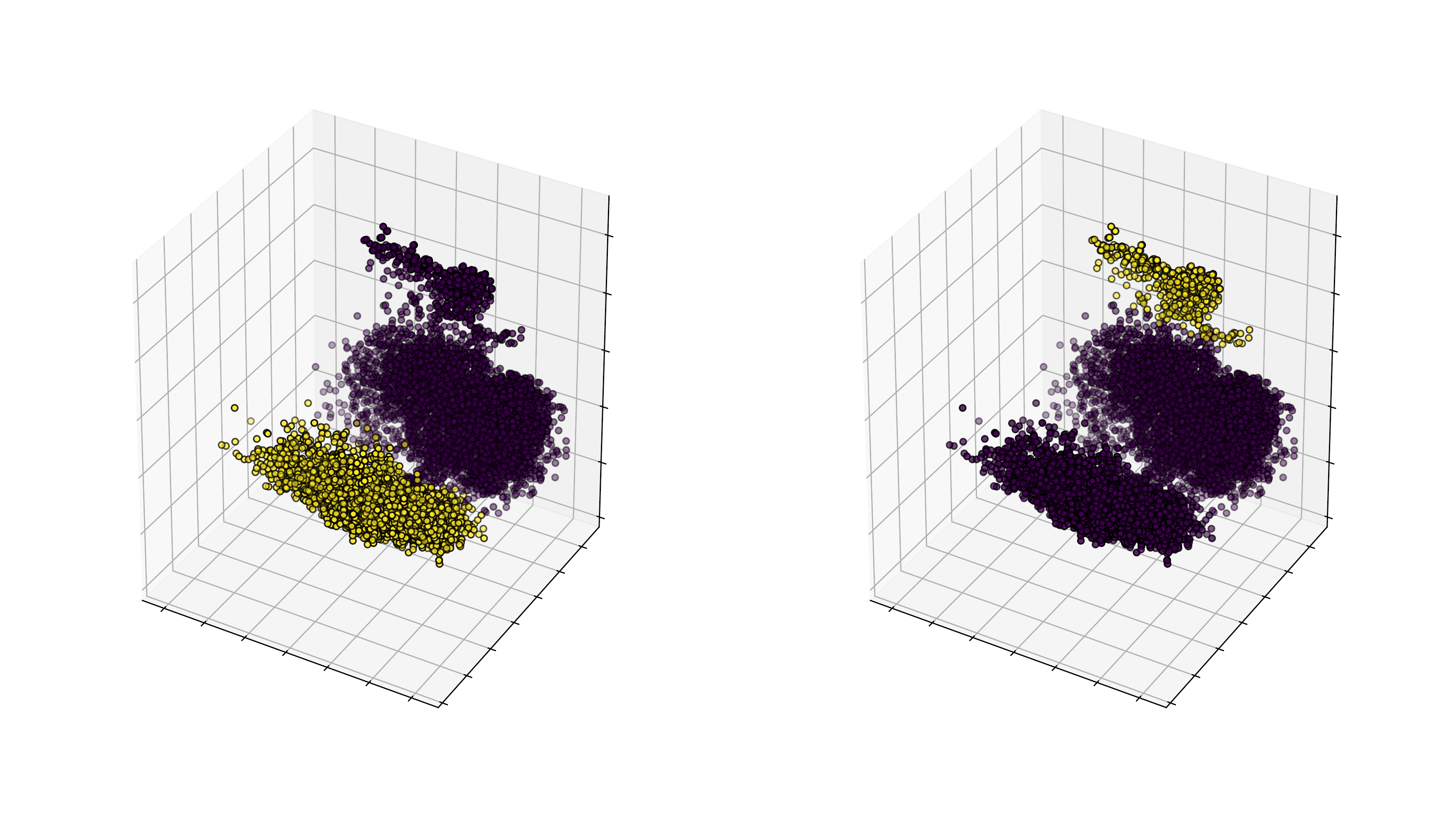}
    \caption{(left) Result of K-means clustering on the ICA of the last-layer
        activations for the target action for our Breakout Trojaned DRL model when $K=2$
        and 10\% of the sample set are poisoned images. (right) The ground
        truth coloring. Clean inputs are purple and the poisoned samples
        populate the yellow cluster.}
    \label{fig:clusters}
\end{figure}

Trojan detection via \textit{activation clustering}~\cite{Chen2018} is a
similar approach that operates on the assumption that Trojaned models select
the target label for different reasons if the trigger is present than if the
input is clean, i.e. the last-layer activations will be different. The
different activation patterns are visible by inspection by performing K-means
clustering on the ICA of the last-layer activations. Since access to the
training data is a prerequisite to this technique, the activation clustering
methodology shares the weakness that it is only applicable under Threat Model
2. Even so, we attempted to use activation clustering on our Targeted-Attacked
Trojan DRL model and found that it was not successful in detecting the Trojan. Our
explanation for activation clustering being insufficient is that we poison
extremely few training samples. The Trojaned model only experienced 20 thousand
poisoned samples across all 80 million training samples (0.025\%) -- with so
few poisoned samples in the training set, the poisoned samples fail to form an
cluster independent of clean samples. Further complicating matters, it appears
that our model appears to select the target action for two different reasons.
In other words, the assumption that the network selects each action for a
single reason when the input is clean does not hold. In fact, even when large
quantities of poisoned samples are present in the training set (10\%), the
poisoned samples are clustered within one of the clean clusters. The poisoned
samples do form their own cluster if $K=3$, although this is something that we
discovered only because we know which samples were poisoned; a realistic
defender would not be able to determine this in practice.

\begin{figure}[ht]
    \centering
    \includegraphics[width=0.99\linewidth]{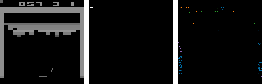}
    \caption{(left) A poisoned state for the Breakout game; the trigger is the $3\times 3$ patch of pixels in the top left corner. (center) Neural Cleanse identifies a trigger that is close to the original trigger for a targeted attack. (right) Neural Cleanse fails to identify the original trigger for the untargeted attack; the four colors are used to illustrate the different triggers identified by Neural Cleanse for each of the four actions in this game.}
    \label{fig:fusion-summary}
\end{figure}

The most promising technique for defending against Trojaned models is Neural Cleanse~\cite{neuralcleanse} as it does not require access to the training data and as
such is applicable even against the strong attackers living under the Threat
Model 1 regime. In Fig.~\ref{fig:fusion-summary} we show the output of Neural
Cleanse on our Targeted-Attacked Breakout model. A defender applying
Neural Cleanse to this model would arguably claim to have detected the attack trigger by
visual inspection. Since the authors of~\cite{neuralcleanse} do not consider untargeted
attacks, i.e. multiple infected labels with single triggers, we
report the expected result that Neural Cleanse is unable to detect the
trigger in our Untargeted-Attacked  model. 

Through our attempts to adapt Trojaned classification network defense methods
to Trojan DRL, we have identified several outstanding issues in the realm
of classification network defense that will in fact carry over to Trojaned DRL networks.
Untargeted attacks are difficult to defend against because untargeted attack triggers induce a distribution over outputs, as shown in Fig.~\ref{fig:climber_actions}, an effect that breaks the assumptions of Neural Cleanse. 
There is no demonstrated
defense for partial Trojans, where the trigger only corrupts a
subset of the output labels. 
Similarly, we have conceptualized
\textit{intra-label} partial Trojans, where only a subset of a specific label
can be corrupted to the target label by the trigger. To the best of
our knowledge, intra-label partial Trojans have not been previously
considered as an outstanding issue by any scholarship.

We also identify challenges unique to Trojaned DRL agents that are not important in
the context of Trojaned classification networks. RL agents are oftentimes used
as controllers in a dynamical system. In the control setting, RL models can be
trained to have (potentially many) continuous control outputs. We have not
demonstrated Trojan attacks in this setting, though we believe it likely that
such agents can be trained with a Trojan that compromises control performance.
This scenario will require entirely new defense techniques as all known
defenses rest on the basis of discrete outputs. Furthermore, we claim that
previous works promising defenses under Threat Model 2 are not effective on
Trojaned DRL agents as large training sets and small amount of 
poisoned inputs inhibit the proper function of such techniques.


\section{Conclusion}
\label{conclusion}
Our work suggests caution in deploying reinforcement learning in high-security safety-critical applications
where the training process is not restricted to a controlled and secure environment. Continuous lifelong reinforcement learning approaches need to be made more resilient to adversarial attack before their deployment to social or mission-critical applications where the learning algorithm cannot be shielded from adversaries.
We have presented a case against outsourced training of DRL agents. Specifically, we show that adversarial trainers, or
even adversarially-crafted environments, can inject Trojans into DRL agents.
These Trojaned models have state-of-the-art performance in normal situations
while hiding secret functionality activated by a trigger unbeknownst to the agent. Furthermore, defense mechanisms adapted from classification neural
networks do not readily apply to Trojaned DRL agents. In future work, we plan to study
Trojan attacks for DRL agents with continuous control outputs.
We also motivate the advancement of defense mechanims,
noting that no existing defense extends to the anticipated vulnerability in DRL
agents with continuous outputs.


\bibliographystyle{splncs04}
\bibliography{./bib/paper,./bib/kacper}



\end{document}